\documentclass [prl,twocolumn,superscriptaddress,showkeys]{revtex4}
\usepackage {bm}
\usepackage {graphicx}
\usepackage {subfigure}
\usepackage {amssymb}
\usepackage {amsmath}
\usepackage {graphicx}
\usepackage {hyperref}

\begin{document}

\title{An assessment of the resolution limitation due to 
radiation-damage in x-ray diffraction microscopy}

\begin{abstract}
X-ray diffraction microscopy (XDM) is a new form of x-ray imaging that
is being practiced at several third-generation synchrotron-radiation
x-ray facilities.  Although only five years have elapsed since the
technique was first introduced, it has made rapid progress in
demonstrating high-resolution three-dimensional imaging and promises
few-nm resolution with much larger samples than can be imaged in the
transmission electron microscope.  Both life- and materials-science
applications of XDM are intended, and it is expected that the
principal limitation to resolution will be radiation damage for life
science and the coherent power of available x-ray sources for material
science.  In this paper we address the question of the role of
radiation damage.  We use a statistical analysis based on the
so-called ``dose fractionation theorem'' of Hegerl and Hoppe to
calculate the dose needed to make an image of a life-science sample by
XDM with a given resolution.  We conclude that the needed dose scales
with the inverse fourth power of the resolution and present
experimental evidence to support this finding.  To determine the
maximum tolerable dose we have assembled a number of data taken from
the literature plus some measurements of our own which cover ranges of
resolution that are not well covered by reports in the literature.
The tentative conclusion of this study is that XDM should be able to
image frozen-hydrated protein samples at a resolution of about 10 nm
with``Rose-criterion'' image quality.
\end{abstract}

\author{M. R. Howells}
\email[Correspondence should be addressed to M. R. Howells, phone 510 486 4949, fax 510 486 7696: ]{mrhowells@lbl.gov}
\author{T. Beetz}
\affiliation{Department of Physics, State University of New York, Stony Brook, NY 11794, USA}
\author{H. N. Chapman} 
\affiliation{Lawrence Livermore National Laboratory, 7000 East Ave., Livermore, CA 94550, USA}
\author{C. Cui}
\affiliation{Advanced Light Source, Lawrence Berkeley National Laboratory, 1 Cyclotron Rd., Berkeley, CA 94720 USA}
\author{J. M. Holton}
\affiliation{Advanced Light Source, Lawrence Berkeley National Laboratory, 1 Cyclotron Rd., Berkeley, CA 94720 USA}
\author{C. J. Jacobsen}
\affiliation{Advanced Light Source, Lawrence Berkeley National Laboratory, 1 Cyclotron Rd., Berkeley, CA 94720 USA}
\affiliation{Department of Physics, State University of New York, Stony Brook, NY 11794, USA}
\author{J. Kirz} 
\affiliation{Advanced Light Source, Lawrence Berkeley National Laboratory, 1 Cyclotron Rd., Berkeley, CA 94720 USA}
\author{E. Lima}
\affiliation{Department of Physics, State University of New York, Stony Brook, NY 11794, USA}

\author{S. Marchesini}
\affiliation{Lawrence Livermore National Laboratory, 7000 East Ave., Livermore, CA 94550, USA}

\author{H. Miao} 
\affiliation{Department of Physics, State University of New York, Stony Brook, NY 11794, USA}

\author{D. Sayre}
\affiliation{Department of Physics, State University of New York, Stony Brook, NY 11794, USA}

\author{D. A. Shapiro} 
\affiliation{Advanced Light Source, Lawrence Berkeley National Laboratory, 1 Cyclotron Rd., Berkeley, CA 94720 USA}
\affiliation{Department of Physics, State University of New York, Stony Brook, NY 11794, USA}

\author{J. C. H. Spence}
\affiliation{Advanced Light Source, Lawrence Berkeley National Laboratory, 1 Cyclotron Rd., Berkeley, CA 94720 USA}
\affiliation{Department of Physics and Astronomy, Arizona State University, Tempe, AZ 85287-1504, USA}

\maketitle

\section{Introduction}

X-ray diffraction microscopy (XDM) is a new form of x-ray imaging that
is now being practiced by the authors at the Advanced Light Source
x-ray facility at Berkeley \cite{Marchesini:2003,Marchesini:2004}.
Similar work has been done at the synchrotron-light sources at
Brookhaven
\cite{Miao:1999,Beetz:2003}, Argonne 
\cite{Robinson:2001,Williams:2003} and the Spring 8 facility in 
Japan \cite{Miao:2002,Miao:2004}. The method works in both 2D and 3D
and can be adapted for both life \cite{Beetz:2003} and materials
sciences. The images are generated in three steps; (\ref{eq1})
diffraction patterns are recorded using coherent x-rays (just one for
2D or a tilt series for 3D) which provides the amplitudes of the
diffracted wave field, (\ref{eq2}) the phases of the wave field are
obtained using variants of phase-retrieval algorithms developed in
other branches of optics
\cite{Gerchberg:1972,Fienup:1978,Elser:2003} and (\ref{eq3}) the 
image is recovered by means of a Fourier transform.

This form of x-ray imaging was first suggested by Sayre
\cite{Sayre:1980} and first demonstrated at Brookhaven in 1999 by Miao, 
Charalambous, Kirz and Sayre \cite{Miao:1999}.  The latter experiment
achieved a resolution of 75 nm using 0.73 keV x-rays and subsequent 2D
experiments have pushed that value down to 7 nm measured in the image
\cite{Miao:2002}. Our own XDM experiments have been done in the energy
region 0.5--0.8 keV while other workers have used energies up to 8
keV. Although all of the above-mentioned groups have achieved 3D
imaging with test objects, the resolution in these experiments is
still several times worse than in 2D. Nevertheless the expansion of
interest in the technique and the progress in developing its
performance has been rapid and we are lead to investigate the
fundamental limits to this form of microscopy. It appears that the
limit for life-science samples will be set by radiation damage, while
for more-radiation-hard materials-science samples it will be set by
the coherent power of available x-ray sources.

In this paper we address the question of the role of damage in setting
a resolution limit to life-science imaging by XDM. This is important
because, XDM is expensive (it needs at least an undulator on a
third-generation synchrotron source) and if it is to have a niche in
which it delivers unique and useful results, then it must produce
performance beyond the limits of other microscopes. In this work, we
refer to the practice of fast freezing the sample and holding it at
low temperature for imaging, as ``cryo-protection''. Such protection
is used, for example, by the Munich group in their
``electron-cryotomography'' system \cite{Medalia:2002} which recently
demonstrated a 3D resolution of 5-6 nm for biological samples of
thickness 0.3-0.6 $\mu $m. Further analysis by the same group
\cite{Plitzko:2002} has indicated that, although the resolution may 
eventually be improved by a factor of 2-3, the thickness is a hard limit 
caused by multiple scattering. Such a thickness limit would not apply to XDM 
so the question becomes: can XDM achieve good enough resolution to produce 
images with similar quality to cryoelectron tomography but of whole cells in 
the 0.5 to say 5-10 $\mu $m size range? A more fundamental reason why the 
issue of resolution is important in these investigations is that the 
resolution achieved by the Munich group is beginning to enable protein 
molecules of known structure in the sample to be recognized. The potential 
for determining the way in which these proteins ``dock'' together and thus 
for throwing light on their function in molecular machines is an exciting 
general goal of these types of ultramicroscopy.

The question of calculating how much dose is \textit{needed} for imaging in a given 
microscope at a given resolution and statistical accuracy is essentially a 
statistical calculation. Such calculations have been presented before for 
x-ray microscopy in general 
\cite{Sayre:1977,Rudolph:1990,Jacobsen:1998} and for XDM 
\cite{Sayre:1995,Marchesini:2003,Shen:2004}. On the other 
hand the question of how much dose can the sample \textit{tolerate} before unacceptable 
degradation occurs to images at a given resolution is not a matter of 
statistics but rather of radiation chemistry and biology. We thus arrive at 
two important quantities that we need to know about in order to estimate the 
dose-limited resolution, the \textit{required dose for imaging} and the \textit{maximum tolerable dose}. Obviously experiments can only be 
successful if the dose employed is greater than the required dose for 
imaging and less than the maximum tolerable dose.

In what follows we will use various techniques to estimate the required dose 
and the maximum tolerable dose. For the required dose we will use an 
estimation method based on the so-called dose fractionation theorem 
\cite{Hegerl:1976,McEwen:1995} which we explain below. To use the 
theorem for a 3D diffraction experiment one needs to know the scattering 
strength of a single voxel.  This cannot normally be 
measured but we will describe simple methods by which it can be calculated 
and will compare the dose-resolution scaling law that results with our own 
XDM measurements. The maximum tolerable dose cannot be estimated by a simple 
calculation so it needs to be inferred from experimental results. We discuss 
below various experiments by ourselves and others that may be able to 
provide information. Since no 3D images of biological samples have yet been 
made by XDM, we try to make the best use of results from other types of 
experiment, 2D XDM, x-ray and electron crystallography and conventional 
electron and x-ray microscopy. Using these methods we will make tentative 
estimates of the future capability of XDM based on the presently-available 
evidence.

\section{The dose fractionation theorem}

The theorem that we will use to simplify our calculation of the required 
dose for imaging was first proved by Hegerl and Hoppe 
\cite{Hegerl:1976}. It states, ``A three-dimensional reconstruction 
requires the same integral dose as a conventional two-dimensional
micrograph provided that the level of (statistical) significance and
the resolution are identical''. The discussion provided by the
originators of the theorem was largely in terms of a single voxel but,
as pointed out by McEwan, Downing and Glaeser \cite{McEwen:1995}, the
conclusion can be immediately generalized to a full 3D object by
recognizing that conventional tomographic reconstructions are
\textit{linear} superpositions of the contributions of the individual
voxels. A similar argument can be used to show that the theorem is
applicable to XDM. McEwan et al also showed by computer simulations
that the validity of the theorem could be extended to include
experimentally realistic conditions of high absorption,
signal-dependent noise, varying contrast and missing angular range.

We consider a single voxel of the type that we would reconstruct in a
fully 3D experiment, which means one with the same width in all three
dimensions.  In order to apply the theorem to predict the ``required
dose for imaging'', we need to know the dose required in an XDM
experiment on the \textit{single voxel alone} for an interesting range
of values of $d$. It would be extremely hard to do such a series of
experiments in practice. However, since the one-voxel experiments are
simple in principle, it is easy to obtain their results by theoretical
analysis which is what we do below. We will study a voxel of size
$d\times d\times d$ which corresponds to correct sampling for
resolving a smallest spatial period of 2$d$ in each coordinate
direction (roughly similar to a Rayleigh resolution of $d$ in standard
microscopy).

To obtain the dose required for the one-voxel experiment we begin by
calculating the x-ray coherent scattering cross section ($\sigma _{s}$)
 of the voxel for scattering into a detector with the right
solid-angle collection to get the resolution $d$. This gives the dose
required to get a given number of x-rays scattered by the voxel into
the detector. The refractive index $\tilde {n} = 1 - \delta - i\beta
$, the intensity absorption coefficient $\mu $, and the
complex electron density $\tilde {\rho }$ that we will need can be
obtained from the tabulated optical constants as described, for
example, in \cite{Kirz:1995} equations 17, 18, 23 and 19 respectively.

\section{Calculation of the coherent scattering cross section of the voxel}
\label{sec:calculation}

Suppose the voxel is of amplitude transparency $T$ surrounded by empty space of 
transparency unity. The Babinet inverse of this scattering object is an 
aperture of transparency $1 - T$ in an opaque screen. Babinet's Principle 
asserts that, outside the (small) area of the incident beam, the diffraction 
pattern of these two objects is the same. The diffracted intensity at 
distance $z $is most easily calculated for the second object 
\cite{Goodman:1968} as follows:
\[
I\left( {x,y} \right) = \frac{I_{\mathrm{in}} \left| {1 - T} 
\right|^2d^4}{\lambda ^2z^2}\mbox{ sinc}^2\left( {\frac{xd}{\lambda z}} 
\right)\mbox{ sinc}^2\left( {\frac{yd}{\lambda z}} \right)\,.
\]

The numerical aperture required to resolve a spatial period 2$d$ is $\lambda 
\mathord{\left/ {\vphantom {\lambda {\left( {2d} \right)}}} \right. 
\kern-\nulldelimiterspace} {\left( {2d} \right)}$ so the full width of the 
detector in both $x$ and $y$ should be $w = {\lambda z}
\mathord{\left/ {\vphantom {{\lambda z} d}}
\right. \kern-\nulldelimiterspace} d$. Thus
\begin{eqnarray}
\nonumber
\frac{\sigma _{s} }{d^2} &=& \frac{\mbox{scattered energy}}{\mbox{incident 
energy}} \cong \frac{I\left( {0,0} \right) w^2}{I_{\mathrm{in}} d^2}\,,\\
\nonumber
&=&\frac{I_{\mathrm{in}} \left| {1 - T} \right|^2d^4}{\lambda ^2z^2}\left( 
{\frac{\lambda z}{d}} \right)^2\frac{1}{I_{\mathrm{in}} d^2} \,,\\
\nonumber
&=& \left| {1 - T} \right|^2\,,
\end{eqnarray}
showing that $\sigma _{s} = \left| {1 - T} \right|^2\,d^2$ which is in
agreement with equation 23 of \cite{Mueller:1976} for example,
as well as being intuitively reasonable. To get the complex
absorbency, $1 - T$ in terms of the material properties of the voxel
we use \cite{Kirz:1995} equation (20) for the wave amplitude $\psi$:
\begin{eqnarray}
\nonumber
 \psi &=& \psi _{0} \,\mbox{e}^{ - {2\pi i \tilde {n}d/ \lambda }} 
 = \psi _{0}\, \mbox{e}^{ - {2\pi i \left( {1 - \beta - i\delta } \right)d/ \lambda }}\,,\\
\nonumber
&=& \psi _{0} \,\mbox{e}^{ - 2\pi i d/\lambda}
\mbox{e}^{ - 2\pi \beta d/\lambda} 
\mbox{e}^{   2\pi i \delta d/\lambda }
\text{, whence} \\ 
\nonumber
 T &=& \mbox{e}^{ - 2\pi \beta d/\lambda}
\mbox{e}^{{2\pi i\delta d/\lambda}}
\simeq
1 - 2\pi d\left( {\beta + i\delta } \right)/\lambda \,,
\end{eqnarray}
where we have introduced the weak-phase-weak-amplitude approximation (which 
will usually be valid for a resolution element which is intrinsically 
small). Recasting this in terms of the complex electron density $\tilde 
{\rho },$ (\cite{Kirz:1995} equation 19), we have:
\[
\left| {1 - T} \right|^2 = {\left( {2\pi d} \right)^2\left| {\beta + i\delta 
} \right|^2} \mathord{\left/ {\vphantom {{\left( {2\pi d} \right)^2\left| 
{\beta + i\delta } \right|^2} {\lambda ^2}}} \right.
\kern-\nulldelimiterspace} {\lambda ^2} = d^2r_{e}^{2} \lambda ^2\left| \rho 
\right|^2\,,
\]
and finally,
\begin{equation}
\label{eq1}
\sigma _{s} = \left| {1 - T} \right|^2\,d^2 = r_{e}^{2} \lambda ^2\left| 
\rho \right|^2d^4\,.
\end{equation}

Thus $\sigma _{s} $ scales as the voxel size to the fourth power. As
we will see this leads to an inverse fourth-power scaling of the dose
with $d$. The scaling with wavelength is dominated by the
lamda-squared term, especially at wavelengths $\ll 2$ nm where the
$\tilde {\rho }$ values of the light elements approach a constant
value.

Equation (\ref{eq1}) is important to our argument and we have checked
it in various ways. Firstly, we took the scattering cross section of a
single electron and summed it coherently over all the electrons in our
voxel. Secondly we used literature calculations of the cross section
of spherical particles of the same size as our voxel
\cite{Henke:1981,London:1989}. The three expressions so
obtained agreed with equation (\ref{eq1}) up to a constant factor of
order one. We may also argue that the contrast ($C$) between the voxel
and vacuum scales as the thickness, i. e. as $d$. The Rose criterion
(see later) says that the number of incident x-rays per unit area,
$N_{0} $, which is proportional to the dose, must satisfy $N_{0} d^2 >
25 \mathord{\left/ {\vphantom {{25} {C^2}}}
\right. \kern-\nulldelimiterspace} {C^2}$. Therefore, since $C$ scales
as $d$, $N_{0} $ scales as $1 \mathord{\left/ {\vphantom {1 {d^4}}}
\right.
\kern-\nulldelimiterspace} {d^4}$.

\section{Relation between flux density and dose}
\label{sec:relation}

\begin{figure}[htb]
\centerline{\includegraphics[width=0.45\textwidth]{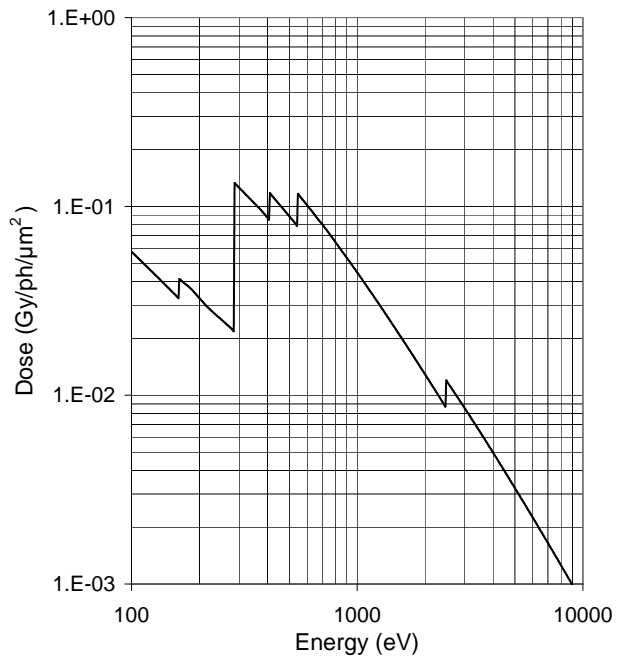}}
\caption{
\label{fig:1}
The surface dose in Gray for an incident x-ray flux density of 
1/$\mu $m$^{2}$ for a range of x-ray energies. The material is taken to be 
protein of empirical formula $\mathrm{H_{50} C_{30} N_{9} O_{10} S_{1}}$
 and density 1.35 gm/cm$^{3}$ }
\end{figure}

Before proceeding to calculate the dose for our case we first make
some definitions and show how the dose is related to the number of
incident particles per unit area. This relationship will be needed in
order to compare published data from different sources. For x-rays of
energy $h\nu $, we know that for any object (with density
\textit{$\varepsilon $}), the number of transmitted x-rays per unit
area $N$ at depth $t$ due to an incident number per unit area $N_{0} $
is given by $N = N_{0} \exp \left( { - \mu t} \right)$ whence the
\textit{energy} deposition per unit volume at the surface, $\left[
\partial \left( {Nh\nu } \right) / {\partial t} \right]_{t = 0} $, is
$\mu \,N_{0} h\nu $. Therefore the dose $D $ (energy deposited per
unit mass) is:
\begin{equation}
\label{eq2}
D = {\mu N_{0} h\nu } \mathord{\left/ {\vphantom {{\mu N_{0} h\nu } 
\varepsilon }} \right. \kern-\nulldelimiterspace} \varepsilon \,.
\end{equation}
$D$ will be in Gray $\left( {\mbox{J} \mathord{\left/ {\vphantom {\mbox{J} 
{\mbox{kg}}}} \right. \kern-\nulldelimiterspace} {\mbox{kg}}} \right)$ if 
the other quantities are in MKS units. The last equation relates the 
incident particle flux density to the dose for given material parameters 
irrespective of $d$. Some numerical values for protein are given in Fig. \ref{fig:1}.

In the case of illumination by an electron beam, the energy deposited per 
unit length of trajectory (and thence per unit volume) is given by the Bethe 
formula, (see for example equation 10.2 of \cite{Reimer:1984}) which we 
have used for some of the entries in Fig. \ref{fig:4}.

\section{Estimation of the dose and calculation of the required imaging dose}

We want to estimate the dose $D$ (Gy) and the number of incident
x-rays per unit area $N_{0} $ required to get a given number $P$ of
x-rays scattered into the detector from the given voxel (The choice
of $P$ will be determined by the statistical accuracy required from
the measurement). The number of photons incident on the voxel is
$N_{0} d^2$ of which a fraction ${\sigma _{s} }
\mathord{\left/ {\vphantom {{\sigma _{s} } {d^2}}} \right. 
\kern-\nulldelimiterspace} {d^2}$ will be scattered into the detector. 
Therefore the requirement is for $N_{0} = P \mathord{\left/ {\vphantom
{P {\sigma _{s} }}} \right. \kern-\nulldelimiterspace} {\sigma _{s} }$
which, from equations (\ref{eq1}) and (\ref{eq2}), leads to:
\begin{equation}
\label{eq3}
D = \frac{\mu \,P\,h\nu \,\,}{\varepsilon \,\sigma _{s} } = \frac{\mu 
\,P\,h\nu }{\varepsilon \,}\mbox{ }\frac{1}{r_{e}^{2} \,\lambda ^2\,\left| 
\rho \right|^2\,d^4}\,,
\end{equation}
and
\begin{equation}
\label{eq4}
N_{0} = \mbox{ }\frac{P}{r_{e}^{2} \,\lambda ^2\,\left| \rho 
\right|^2\,d^4}\,.
\end{equation}

\begin{figure}[htb]
\subfigure[]
	{
	 \label{fig:2a}
	 \includegraphics[width=0.45\textwidth]{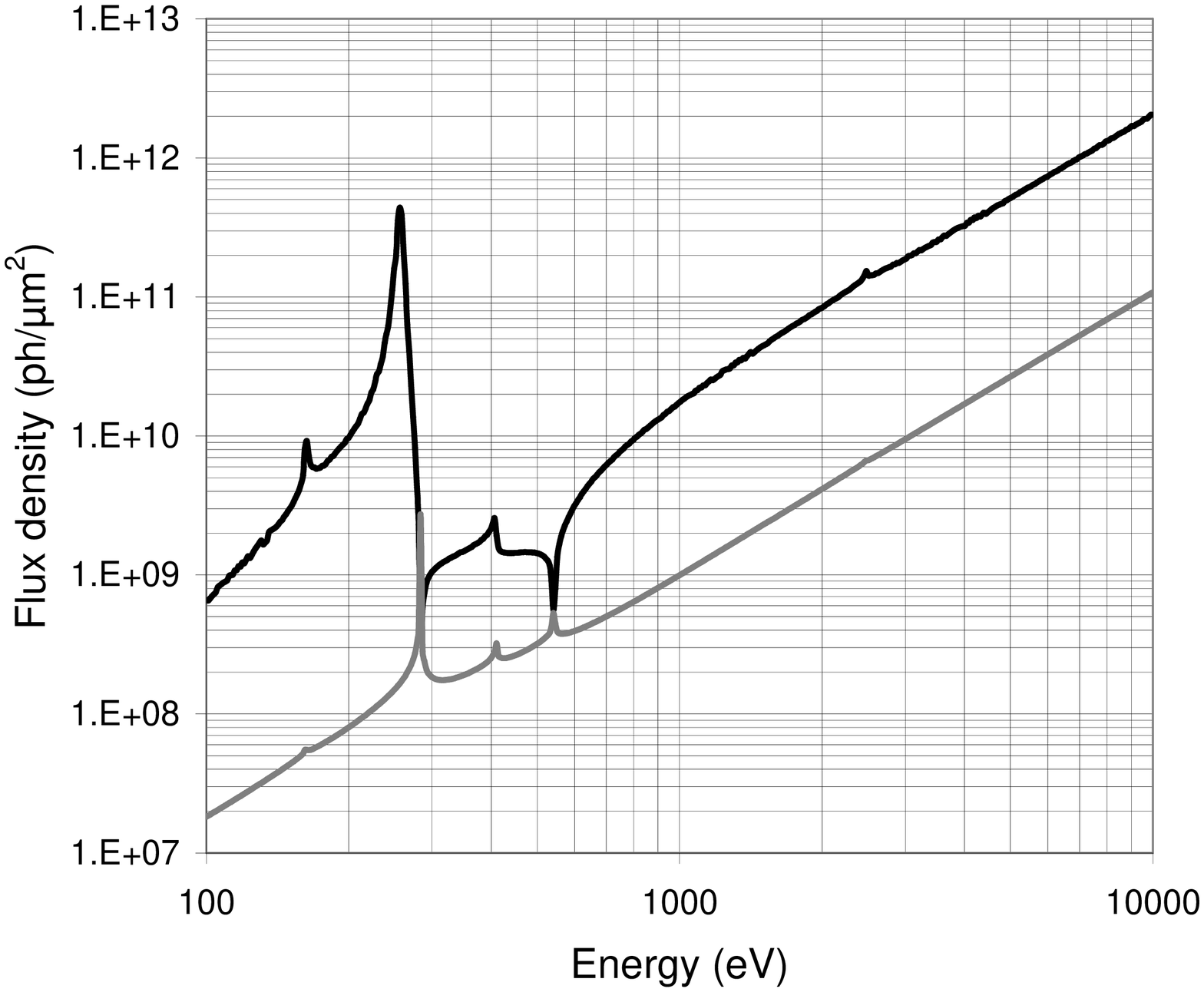}
	}\\
\subfigure[]
	{
	\label{fig:2b}
	\includegraphics[width=0.45\textwidth]{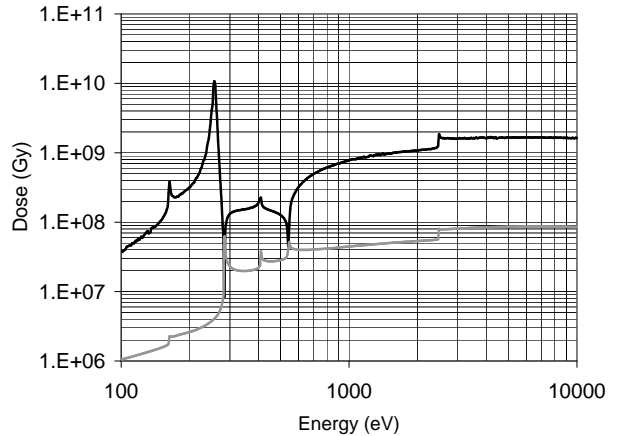}
	}
\caption{(a) The flux density and (b) the dose required to 
visualize a 10 nm cubic voxel of protein of empirical formula
$\mathrm{H_{50} C_{30} N_{9} O_{10} S_{1} }$ and density 1.35
gm/cm$^{3}$ against a background of water (black) and vacuum (gray)
according to the Rose criterion.}
\label{fig:2}
\end{figure}
As examples we show the flux and dose curves (Fig. \ref{fig:2}) for
a protein sample of empirical formula $\mbox{H}_{\mbox{50}} C_{30}
N_{9} O_{10} S_{1} $ and density 1.35 gm/cm$^{3}$ as a function of
x-ray energy.  The curves are for a voxel size (resolution) of 10 nm
and statistical accuracy based on the Rose criterion
\cite{Rose:1948}. The latter is an experimentally-based criterion for
reliable detectability of a feature against background noise. The
requirement is generally that the feature signal should be five times
greater than the rms background noise. When the noise is the shot
noise of the feature signal itself then it is conventional to set the
particle count equal to 25. The flux curve (Fig. \ref{fig:2a}) is dominated
by the $\lambda ^{ - 2}$ scaling of the cross section. This argues for
using the longest possible wavelength for these experiments. On the
other hand 
the wavelength should be shorter than, say, a quarter to a half of the 
resolution so that the diffraction angle is not too large, and short enough 
that the sample is a weak absorber ($<20{\%}$, say), so that data analysis can 
proceed on the basis of the Born approximation. Unlike the flux, the dose 
does not show strong energy dependence above about 1 keV. This is because 
the roughly $\lambda ^{5 \mathord{\left/ {\vphantom {5 2}} \right. 
\kern-\nulldelimiterspace} 2}$ scaling of the absorption coefficient tends 
to cancel the wavelength dependence of ${h\nu } \mathord{\left/ {\vphantom 
{{h\nu } {\sigma _{s} }}} \right. \kern-\nulldelimiterspace} {\sigma _{s} }$ 
in equation (\ref{eq3}).

Equation (\ref{eq3}) also allows the calculation of the ``required
imaging dose'' as a function of resolution $d$. We have evaluated that
for protein against a background of water for 1 keV and 10 keV as
shown by the solid and dashed straight lines in Fig. \ref{fig:4}. One can see
that the change in dose from 1 keV to 10 keV is not very significant.

Also from (\ref{eq3}) the resolution scaling of the dose is seen to be $1 
\mathord{\left/ {\vphantom {1 {d^4}}} \right. \kern-\nulldelimiterspace} 
{d^4}$ and is determined entirely by the cross
section$\mathfrak{u}\mathfrak{g}$ow applying the dose fractionation
theorem we may say that the \textit{same dose }will be required to
measure the same $d\times d\times d$ voxel to the same statistical
accuracy in a 3D tomography experiment. Hence, the
inverse-fourth-power scaling with $d,$ will also apply to a 3D sample.

It is important to note that the flux, predicted by Fig. \ref{fig:2}
to be required for a 10 nm XDM experiment on protein against a
background of water, needs to be delivered to the sample as a coherent
x-ray beam. Part of the attraction of XDM is that such coherent x-ray
beams are already available from undulators on current
synchrotron-radiation sources such as the Advanced Light Source at
Berkeley USA. For example our present experiments are done using
exposure times of several tens of seconds per view in a tomographic
tilt series using a general-purpose beam line. The use of a
purpose-designed beam line plus the soon-to-be-completed ALS upgrade
would improve that by a factor of about a thousand.

\section{Measurements of the required imaging dose}
\label{sec:measurements}

In order to test the calculations of the required imaging dose we have
carried out the following series of measurements. Two-dimensional
diffraction patterns were recorded using a series of exposure times
that increased on a logarithmic grid. The patterns were analyzed by
first taking an azimuthal average so as to produce a relation between
diffracted signal and spatial frequency. A cut-off frequency was then
determined from where the diffracted signal reached the noise floor of
the detector. The exposure times were converted to dose units and the
cut-off frequencies to spatial half-periods giving a relation between
dose and resolution. One of these relationships, taken using
freeze-dried yeast, is plotted in Fig. \ref{fig:4} (crosses). It shows
that the predicted magnitude is roughly right (remember that the plot
is for frozen-hydrated material so the precise agreement shown should
not be taken too seriously) and the inverse-fourth-power scaling with
resolution is well reproduced. Another noteworthy feature is that the
plot follows a good straight line on the log-log plot all the way up
to the maximum dose employed. We interpret that to mean that the
resolution was not compromised by damage up to the maximum integrated
dose employed.  Furthermore we suggest that the eventual departure
from straightness of such plots may be a good indicator of the onset
of a loss of resolution due to damage. Although the data described
above and plotted in Fig. \ref{fig:4} are typical of the majority of
the data we have taken, we wish to point out that some of our data
have shown scaling laws that departed significantly from
inverse-fourth-power, the lowest power so far being -3.1.

\section{Measurement of the maximum tolerable dose}

Since the ninteen seventies \cite{Glaeser:1971}, there has been strong 
interest in understanding the role of radiation damage in various forms of 
imaging of life-science samples. This has been important in direct imaging 
by electron and x-ray microscopes and in reconstructive imaging by methods 
such as the single-particle technique \cite{Franck:1996} and by x-ray and 
electron crystallography. During this time there has been a continual growth 
in the power of electron and x-ray sources and an interest in using smaller 
crystals in crystallography and larger numbers of images in single-particle 
work, all of which has generated a motivation to push data collection to the 
limits allowed by damage. Thus radiation damage studies are still very much 
on the current agenda (see for example, the comments of Henderson 
\cite{Henderson:2004} and the review of a recent workshop at Argonne, USA 
on the subject by Garman and Nave \cite{Garman:2002}). Our task to 
judiciously apply the latest information from these studies to the issue of 
determining the maximum tolerable dose for XDM.

\begin{figure}[htb]
\subfigure[]
	{
	\includegraphics[width=0.45\textwidth]{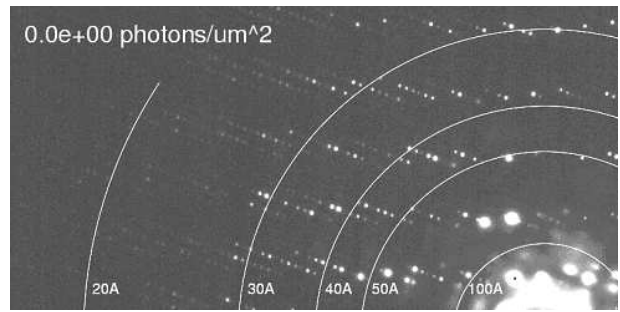}
	\label{fig3a}
	}\\
\subfigure[]
	{
	\includegraphics[width=0.45\textwidth]{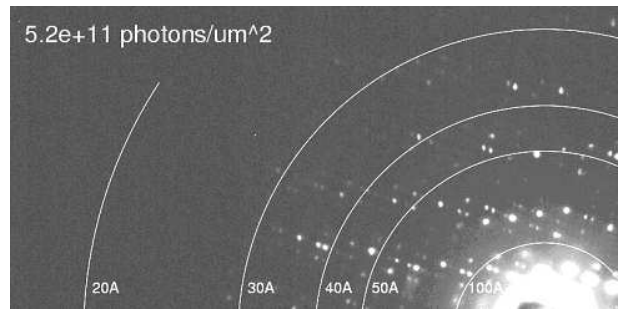}
	\label{fig3b}
	}\\
\subfigure[]
	{
	\includegraphics[width=0.45\textwidth]{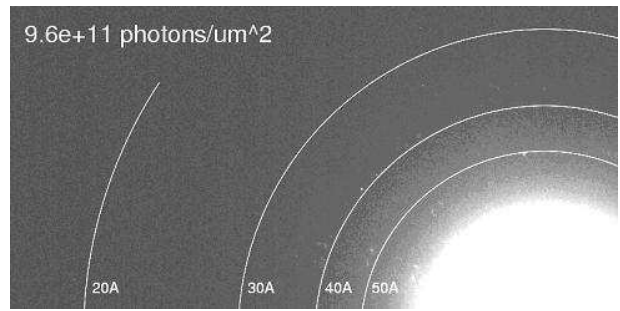}
	\label{fig3c}
	}
\caption{ Three spot patterns from the series described in the text recorded 
from the ribosome crystal. Many of the spots seen at the beginning of the 
sequence (a) have faded by the middle (b) and essentially all are gone by 
the end (c). The full sequence can be seen as a movie at 
\href{http://bl831.als.lbl.gov/~jamesh/ribo_blast/diffraction.gif}
{http://bl831.als.lbl.gov/$\sim $jamesh/ribo{\_}blast/diffraction.gif}
}
\label{fig:3}
\end{figure}

\begin{figure}[htb]
\centerline{\includegraphics[width=0.45\textwidth]{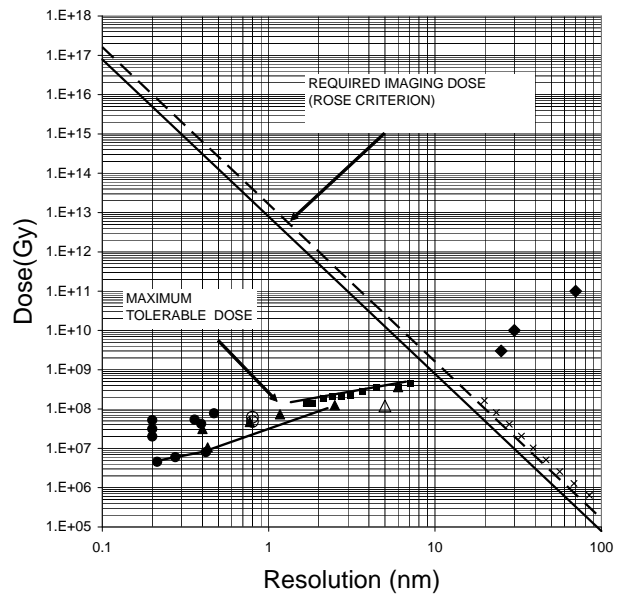}}
\label{fig4}
\caption{graph summarizing information on the required imaging dose and the 
maximum tolerable dose. The reason why experiments on crystals are
seen to be done successfully with around $10^8$ times less than the
required imaging dose is that the dose applied to a crystal is shared
among at least that number of copies of the unknown object. The
required imaging dose is calculated for protein of empirical formula
$\mathrm{H_{50} C_{30} N_{9} O_{10} S_{1}}$ and density 1.35
gm/cm$^{3}$ against a background of water for x-ray energies of 1 keV
(solid line) and 10 keV (dashed line).  Some of our measurements of
imaging dose are plotted as crosses (see text).  The maximum tolerable
dose is obtained from a variety of experiments by ourselves and from
the literature as described in Table \ref{tab:1}. The types of data
from the literature are identified by the symbols as follows: filled
circles: x-ray crystallography, filled triangles: electron
crystallography, open circles: single-particle reconstruction, open
triangles: electron tomography, diamonds: soft x-ray microscopy
(including XDM), filled squares: ribosome experiment (see text).}
\label{fig:4}
\end{figure}

Our resolution goal in life-science XDM is set by the considerations 
discussed in the introduction to be 3-10 nm. This does not correspond to the 
resolution goals in x-ray or electron crystallography where much higher 
resolution levels of 0.15-0.3 nm, that can lead to atomic-resolution 
structure determinations, are usually desired. In fact a significant part of 
the x-ray damage literature refers to \textit{primary} damage; that is damage to the highest 
resolution structures. Nevertheless we have tried to find reports in the 
crystallography literature that have at least some reference to damage at 
resolution values closer to our range of interest and also give quantitative 
spot-fading data. We have also added information from imaging methods; 
electron tomography and single-particle measurements as well as a few 
results from XDM and x-ray microscope experiments to the compilation in 
Table \ref{tab:1} and Fig. \ref{fig:4}.

Although the data from the literature noted above give quite a consistent 
picture as between electron and x-ray measurements (as noted by Henderson 
\cite{Henderson:1990}), the x-ray measurements have a large gap in the 
resolution region of principal interest to us. This lead us to carry
out spot-fading measurements ourselves using beam line 8.3.1 at the
Advanced Light Source at Berkeley. The experiments were done by
J. M. Holton using the established crystallography facilities of the
beam line with a ribosome crystal grown by Prof. J. Cate of University
of California Berkeley. The total exposure at 10 keV x-ray energy was
about 24 hours with high-dose-rate (wide-slit) exposures to do damage
alternated with low-dose (narrow-slit) exposures to measure the
spots. The spot patterns at the beginning, middle and end of this
sequence are shown in Fig. \ref{fig:3} and the whole process can be
seen as a movie at 
\href{http://bl831.als.lbl.gov/~jamesh/ribo_blast/diffraction.gif}
{http://bl831.als.lbl.gov/$\sim $jamesh/ribo{\_}blast/diffraction.gif}.
 The following points can be noted.
\begin{itemize}
\item The crystal has a unit cell size a=b=693 {\AA}, c=1388 {\AA} with space 
group I4122 and it diffracted out to about 10{\AA} when it was undamaged.

\item As the dose increased, the intensity of the Bragg spots faded without 
increase of the spot size starting from the highest-resolution spots.

\item As the intensity in the (high-angle) Bragg spots diminished, that in the 
central (small-angle) pattern increased strongly.

\item The number and resolution of the spots which faded for each increment of the 
dose was quantified by the DISTYL software \cite{Zhang:2004} as listed
in Table \ref{tab:1} and Fig. \ref{fig:4}.

\item As shown in Fig. \ref{fig:4} the new results are consistent with the
earlier ones and, taken together, the data in the resolution range
0.1--10 nm suggest an approximate straight line on the log-log plot
with slope corresponding roughly to the linear relationship:
$\mbox{dose(Gy) = 10}^8\times \mbox{resolution(nm)}$.  The data in
this region are all from crystallography (electron and x-ray) and
electron imaging.
\end{itemize}

We now turn to the remaining data above resolution 10 nm. There are
only three data points, all coming from soft x-ray imaging
experiments. Two of them differ from the other data in the sense that
they represent experiments in which damage effects, meaning changes to
the image with increasing dose, were not seen. These therefore
represent only a lower bound on the maximum tolerable dose. In the
third (the furthest to the right in Fig. \ref{fig:4}) damage was not
seen at $10^{10}$Gy but was seen at $5\times 10^{11}$Gy so that
experiment did reach an end point.

\section{The meaning of data summarized in Fig. \ref{fig:4}}

The data summarized in Fig. \ref{fig:4} refer to two different dose
levels; the required dose for imaging (the continuous and dashed black
lines), and the maximum tolerable dose (the mostly isolated points
forming a rough straight line going uphill to the right). On the
left-hand side of the crossover of the two lines it is obvious that
the required dose for imaging (by XDM) is far greater than the maximum
tolerable dose. This reflects the fact that XDM experiments cannot be
done at those resolution values. Experiments that share the dose over
multiple copies of the sample (such as crystallography) have a major
dose advantage and, with enough copies, such experiments can be done
successfully up to the resolutions and doses indicated in Table
\ref{tab:1} and Fig. \ref{fig:4}. Obviously the dose advantage factor
of crystallography compared to XDM, in which latter only a single copy
of the unknown object is available, will be related to the number of
copies of the unknown object contained in the crystal(s). However, one
would not expect the relationship to be simple because the degree of
coherent enhancement would depend in a complicated way on the
experimental geometry.

Based on these understandings, the region of the graph where XDM experiments 
could be expected to be successful is the triangular region between the two 
lines to the right of the crossover. The best resolution in the ``good'' 
region is evidently at the crossover of the two ``lines''. We see that this 
is at around 10 nm.

\section{Discussion}
\label{sec:discussion}

Initial studies \cite{Glaeser:1978,Hayward:1979} showed that 
cryo-protection was quite successful in electron microscopy and 
crystallography and a similar technique was later adopted as a standard 
method in x-ray crystallography \cite{Henderson:1990} where it was even 
more successful. The idea to extend cryo-protection to lower temperatures 
including liquid helium temperatures \cite{Dubochet:1981} has also been 
around for some time in the electron-microscopy community and there is some 
consensus that modest benefits may be obtained. The situation in the x-ray 
crystallography community is still much more uncertain. The Argonne workshop 
cited above had several papers on the subject without reaching a clear 
conclusion \cite{Garman:2002}. On the other hand, a recent paper by 
Teng and Moffat \cite{Teng:2002}assessing the range 40-150\r{ }K provided 
convincing evidence that temperatures below 150\r{ }K provide no improvement 
in the dose limit for \textit{primary} radiation damage. However, they did find improvement 
in the dose limit for secondary and tertiary damage which is in a 
poorer-resolution range than primary damage. This suggests that helium 
temperatures might be useful for x-ray experiments in our resolution range. 
Although the available (non-XDM) evidence is not unanimous on this point, we 
would like to explore it experimentally for XDM. However, for the moment we 
are only equiped for liquid-nitrogen temperature.

The work that we are doing consists of an active program of studies of yeast 
cells as a model sample and it is one of our goals to determine the maximum 
tolerable dose in that context. We are also developing the technique of 
measuring and reconstructing tilt series using radiation-hard samples of, 
for example, 50-nm gold balls \cite{Marchesini:2003,Marchesini:2004}. 
At the time of writing we have recorded diffraction patterns at 750 eV of 
freeze dried yeast and have reconstructed XDM images in 2D and stereo pair 
but not 3D from them. The reconstructed resolution currently achieved with 
freeze-dried yeast is 50 nm with a dose level of $2\times 10^7$Gy. With 
frozen-hydrated yeast we have 520 eV diffraction patterns out to 25 nm 
resolution but not yet a reconstruction. The freeze-dried samples were 
overexposed to look for signs of radiation damage. No degradation of the 
resolution by radiation damage was seen in this experiment although some 
shrinkage was observed. The shrinkage probably was a radiation effect 
\cite{Berriman:1986} but we have not seen such shrinkage in frozen 
hydrated samples which are our principal interest. Although the yeast study 
is obviously not finished, the present paper is being submitted now in the 
spirit of a progress report to meet the deadline for the special issue of 
the journal.

\section{Conclusion}

Our work in this area has been directed toward understanding the
resolution limit set by radiation damage in the imaging of
frozen-hydrated samples using XDM The experimental evidence we have
presented here suggests that, if the maximum tolerable dose for XDM is
similar to that for the other methodologies represented in Table
\ref{tab:1} and Fig. \ref{fig:4}, then we should be able to get to 10
nm resolution with ``Rose-criterion'' image quality. This is a
prediction based on the assumptions noted, not a demonstration. At the
present time this is as far as the data allow us to go in predicting
the future capability of XDM. However, we believe that in the
reasonably near future we will have further experimental evidence to
report.

\begin{acknowledgments}
The authors are grateful to Prof. J. Cate for permission to use the ribosome 
crystal, to Prof. R. M. Glaeser for extended and valuable discussions and 
comments and to Dr. H. A. Padmore for sustained encouragement of this work. 
The Lawrence Berkeley National Laboratory authors and the Advanced Light 
source facility at Lawrence Berkeley National Laboratory are supported by 
the Director, Office of Energy Research, Office of Basics Energy Sciences, 
Materials Sciences Division of the U. S. Department of Energy, under 
Contract No. DE-AC03-76SF00098. The work of the LLNL authors was performed 
under the auspices of the U.S. Department of Energy by University of 
California, Lawrence Livermore National Laboratory under Contract 
W-7405-Eng-48. The Stony Brook group has been supported by National 
Institutes of Health grant number 1R01 GM64846-01, and by U. S. Department 
of Energy grant number DEFG0204ER46128.
\end{acknowledgments}

\begin{table*}[htbp]
\begin{tabular}
{|l|l|l|l|l|l|l|l|l|}
\hline
Res'n& Dose & Experiment & Particle energy& Reference& Sample: (crystal \\
(nm)& (Gy)&  & (keV)& &  where stated) \\
\hline
\multicolumn{6}{|l|}{Electrons}
  \\
\hline
0.43& 1.06E+07& spot fading & 100& \cite{Glaeser:1978}& catalase \\
2.5& 1.25E+08& spot fading & 100& \cite{Glaeser:1978}& catalase \\
5.0& 1.20E+08& tomography& 300& \cite{Medalia:2002,Plitzko:2002}& cell in amorphous ice \\
0.77& 4.67E+07& spot fading& 100& \cite{Hayward:1979}& purple membrane \\
1.17& 7.35E+07& spot fading& 100& \cite{Hayward:1979}& purple membrane \\
0.4& 3.12E+07& spot fading& 100& \cite{Hayward:1979}& purple membrane \\
0.8& 4.80E+07& single particle reconst'n& 100 & \cite{Glaeser:2004}& 
protein single molecules\\
0.8& 6.20E+07& single particle reconst'n& 100& \cite{Glaeser:2004}& protein single molecules \\
\hline
\multicolumn{6}{|l|}{X-rays} \\
\hline
30& 1.00E+10& microscopy Berlin& 0.52& \cite{Schneider:1998} & cell in amorphous ice \\
60& 5.00E+11& microscopy Brookhaven& 0.52& \cite{Maser:2000}& cell in amorphous ice\\
0.2& 2.00E+07& generic limit & 8-12& \cite{Henderson:1990} & organic material \\
0.2& 3.10E+07& spot fading& 13.1& \cite{Burmeister:2000} & myrosinase \\
0.36& 5.40E+07& spot fading& 12.4& \cite{Sliz:2003}& various\\
0.47& 7.80E+07& spot fading & 12.4& \cite{Sliz:2003} & various \\
0.39& 4.20E+07& spot fading & 12.4& \cite{Sliz:2003} & various \\
25.0& 3.00E+09& XDM Berkeley & 0.52& this work & yeast cell freeze dried \\
0.42& 8.00E+06& spot fading& 11& \cite{Glaeser:2000}& bacteriorhodopsin \\
0.28& 5.95E+06& spot fading& 11& \cite{Glaeser:2000}& bacteriorhodopsin \\
0.21& 4.55E+06& spot fading& 11& \cite{Glaeser:2000}& bacteriorhodopsin \\
7.1& 4.44E+08& spot fading& 10& this work & ribosome \\
6.0& 3.64E+08& spot fading& 10& this work & ribosome \\
4.5& 3.49E+08& spot fading& 10& this work & ribosome \\
3.7& 2.85E+08& spot fading& 10& this work & ribosome \\
3.1& 2.22E+08& spot fading& 10& this work & ribosome \\
2.7& 2.14E+08& spot fading& 10& this work & ribosome \\
2.4& 2.06E+08& spot fading& 10& this work & ribosome \\
2.1& 1.90E+08& spot fading& 10& this work & ribosome \\
1.8& 1.43E+08& spot fading& 10& this work & ribosome \\
1.7& 1.43E+08& spot fading& 10& this work & ribosome \\
\hline
\end{tabular}
\caption{data types and sources used to estimate the maximum 
tolerable dose}
\label{tab:1}
\end{table*}

\end{document}